\documentclass[twoside,twocolumn]{photon2007}
\usepackage[latin1]{inputenc}
\usepackage{wrapfig,rotating}
\usepackage{amssymb,amsmath,array}

\newcommand{\etal}{{\it et al.}}
\newcommand{\ep}{\varepsilon}

\newcommand{\eqsal}[1]{\begin{align}#1\end{align}}
\newcommand{\cf}[1]{{Fig.~\ref{#1}}}

\begin{document}
\title{
%%%%   Paper title goes here  %%%%%%%%%%%%%%
Transition Distribution Amplitudes for $\gamma^\star \gamma$ collisions} 
%% 
%***********************************************************************
% AUTHORS INFORMATION AREA
%***********************************************************************
\author{
J.P. Lansberg$^{1,2}$, B. Pire$^{2}$ and L. Szymanowski$^{2,3}$
% DO NOT MODIFY THE FOLLOWING '\vspace' ARGUMENT
\thanks{L.Sz. is supported in part by the Polish Grant 1 P03B 028 28 and by the FNRS (Belgium).}
\vspace{.3cm}\\
{\it $^1$ Institut f\"ur Theoretische Physik, Universit\"at Heidelberg, D-69120 Heidelberg, Germany}
\vspace{.1cm}\\
{\it $^2$  
CPhT, \'Ecole Polytechnique CNRS, 91128 Palaiseau, France} 
\vspace{.1cm}\\
{\it $^3$  Soltan Institute for Nuclear Studies, Warsaw, Poland and }\\
{\it University of Li{\`e}ge, Li{\`e}ge, Belgium}\\
}

%%***********************************************************************
% END OF AUTHORS INFORMATION AREA
%***********************************************************************

\maketitle

\begin{abstract}
We  study  the exclusive production of $\pi\pi$  and $\rho\pi$ 
in hard $\gamma^\star \gamma$ scattering in the forward 
kinematical region where the virtuality of one photon provides us with a hard scale in the 
process. 
The newly  introduced concept of Transition Distribution Amplitudes  (TDA) is used to
 perform a QCD calculation of these reactions thanks to two simple models for TDAs. 
The sizable cross sections for  $ \rho \pi$ and $\pi \pi $  production 
 may be tested at intense electron-positron colliders such as CLEO and $B$ factories (Belle and \sc{BaBar}). 
\end{abstract}

\section{Introduction}
In a series of recent papers \cite{PStot}, we have advocated that  factorisation theorems \cite{fact} 
for exclusive processes may be extended to the case of other reactions such as
$\pi^-\,\pi^+ \, \to \,\gamma^\star\,\gamma$ and $\gamma_L^\star\,\gamma \to \, A B$
in the kinematical regime where the virtual photon is highly virtual (of 
the order of the energy squared of the reaction) but the momentum transfer $t$
is small. This enlarges the successful description of deep-exclusive $\gamma \gamma$
reactions  in  terms   of  distribution  amplitudes \cite{ERBL}  and/or
generalised  distribution amplitudes \cite{GDAAPT}  on the  one  side and
perturbatively   calculable  coefficient  functions   describing  hard
scattering at  the partonic  level on the  other side.  The  reactions 
   \begin{equation} 
  \gamma_L^\star\,\gamma \to  \rho^\pm\pi^\mp,~\gamma_L^\star\,~\gamma \to  \pi^\pm\pi^\mp,~ 
\gamma_L^\star\gamma \to \pi^0\pi^0, \nonumber
  \label{gagapi}
\end{equation} 
in the  near forward region and for large virtual photon invariant mass 
$Q$,  may  be studied in detail at intense electron colliders such as
 those which are mostly used as $B$ factories.

 With the kinematics  described in \cf{fig:ggstarAB}, we define the $\gamma \to \pi$ transition distribution 
amplitudes (TDAs)  $T(x, \xi, t)$ as the Fourier transform of matrix elements 
$\langle     \pi(p_\pi)|\, {\cal O} \,|\gamma(p_\gamma) \rangle$ where 
${\cal O} = \bar{\psi}(\frac{-z}{2})[\frac{-z}{2},\frac{z}{2}]\,\Gamma{\psi}(\frac{z}{2})$
 with $\Gamma = \gamma^\mu, \gamma^\mu \gamma^5,
\sigma^{\mu\,\nu}$.  The Wilson line $[\frac{-z}{2},\frac{z}{2}]$  provides  the QCD-gauge
invariance for  non-local operators  and equals unity in  a light-like
(axial) gauge. We do not write the 
 electromagnetic Wilson line caused by the presence of the photon, since 
we choose an electromagnetic axial gauge for the photon.
We then factorise the amplitude of the process  $  \gamma_L^\star \gamma \to A \pi$
as
\begin{equation}
\label{} 
%{\cal  M} (Q^2, \xi, t)\propto  
\int dx dz \,\Phi_{A}(z)  M_{h}(z,x,\xi) T(x, \xi, t)\;,
\end{equation}
 with a hard amplitude $M_{h}(z,x, \xi)$   
and  $\Phi_{A}(z)$   is  the   hadron  $A$ distribution amplitude (DA). 

\begin{figure}[h]
\centering{\includegraphics[width=4cm]{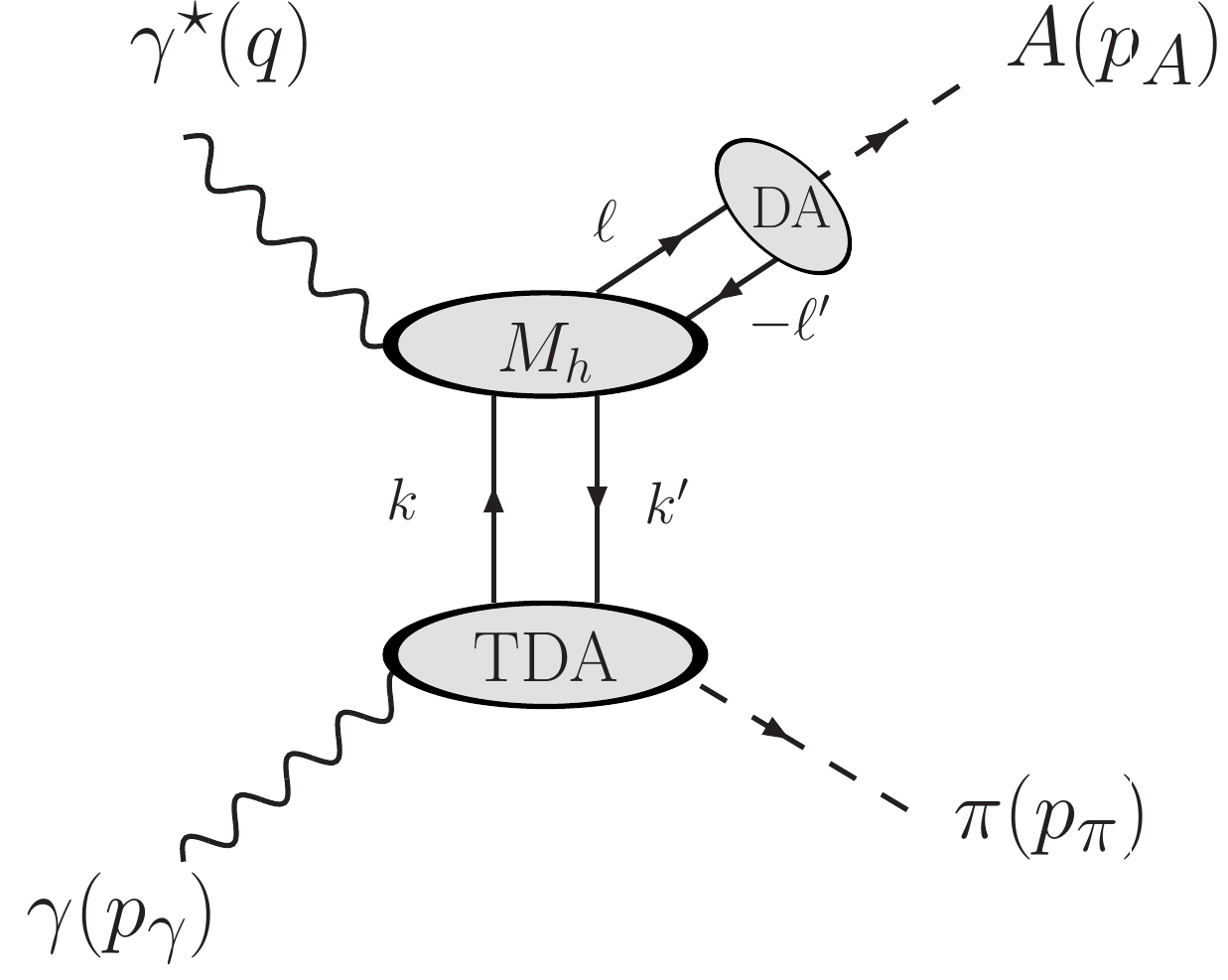}}
\caption{The factorised amplitude for $\gamma^\star \gamma \to A \pi$ at small transfer
momentum.}
\label{fig:ggstarAB}
\end{figure}

The variable $z$ is as usual the light-cone momentum fraction carried by the quark entering the
meson $A$, $x+\xi$ (resp. $x-\xi$) is the corresponding one for the quark leaving (resp. entering) 
the TDA. The skewness variable $\xi$ describes the loss of light-cone momentum of the incident 
photon and is connected to the Bjorken variable $x_B$.

Contrarily to the case of generalised parton distributions (GPD) 
where the forward limit is related to the conventional parton distributions 
measured in the deep inelastic scattering (DIS), there is no such 
interesting constraints for the  new TDAs.
The only constraints are sum rules obtained by taking the local limit of
the corresponding operators and possibly soft limits when the momentum
of the meson in the TDA vanishes. Lacking any non-perturbative calculations of
matrix element defining TDAs, we are forced to build toy models to get 
 estimates for the cross sections, to be compared with
future experimental data. 

For definiteness, let us consider the $ \gamma  \to\pi^{-} $ vector
TDAs which is given by ($P=\frac{p_{\pi^-}+p_\gamma}{2}$, 
$\Delta=p_{\pi^-}-p_\gamma$):

\eqsal{
 \int \frac{d z^-}{2\pi}&e^{ix P^+ z^-} 
\langle  \pi^-| \bar{d}(\frac{-z}{2})\Big[\frac{-z}{2};\frac{z}{2}\Big]\gamma^\mu{u}(\frac{z}{2})|
\gamma \rangle  \nonumber\\ =
&\frac{1}{P^+} \frac{i\;e}{f_\pi}\epsilon^{\mu \ep P \Delta_{\perp}} 
V^{\pi^-}(x,\xi,t)\; . \nonumber
}

 A sum rule  may  be derived  for  this  photon to meson
TDA.  Since the local matrix element  appears in radiative 
weak decays, we can relate it to the 
form factor $F_V$ : 
\begin{equation} 
\int^1_0 dx \, V^{\pi^{\pm}}(x,\xi,t)= \frac{f_\pi}{m_\pi} \,F^{\pi^{\pm}}_V(t), 
\label{srAV}
\end{equation}
with $F_V^{\pi^{\pm}}=0.017  \pm 0.008$.

Let us consider the $\rho_L^\pm\,\pi^\mp$ production case  when the $\rho$ flies 
in the direction of the virtual photon and the $\pi$ enters the TDA. Note that
 in the neutral $\rho^0 \,\pi^0$  case, the TDA process 
is forbidden by $C$-conjugation.

\noindent For definiteness, we choose, in the CMS of the meson pair,  
$p=\frac{Q^2+W^2}{2(1+\xi)W}(1,0,0,-1)$ 
and $n=\frac{(1+\xi)W}{2(Q^2+W^2)}(1,0,0,1)$  and we express the momenta 
 trough a Sudakov decomposition :

\begin{center}
\begin{tabular}{ll}
$p_\gamma= (1+\xi) p$, &\\
$q= \frac{Q^2+W^2}{1+\xi} n - \frac{Q^2}{Q^2+W^2}(1+\xi) p$,\\
$p_{\pi^-}=(1-\xi) p - \frac{\Delta_T^2}{1-\xi} n + \Delta_T$,\\
\end{tabular}
\end{center}
with $ \Delta_T^2=\frac{1-\xi}{1+\xi} t$.

We can see that $\xi$ is determined by the external kinematics 
through $\xi\simeq\frac{Q^2}{Q^2 + 2W^2}$ -- similarly to $x_B=\frac{Q^2}{Q^2 + W^2}$ to which it
is linked via the simple relation $\xi\simeq\frac{x_B}{2-x_B}$.

The amplitude ${\cal  M}^{TDA}_{\gamma^\star \gamma} (Q^2, \xi,t)$ for  the reaction $\gamma_L^\star \gamma \to \rho_L^+ \,\pi^- $  
is  proportional to $V(x, \xi, t)$; it reads
\begin{equation}
\label{amprho} 
 -\int_{-1}^1 dx \int_{0}^1dz\frac{f_\rho}{f_\pi} \phi_\rho(z)
  M_{h}(z,x,\xi) V(x, \xi, t)\;,\nonumber
\end{equation}
 where the hard amplitude $M_{h}(z,x,\xi)$ is ($\bar  z = 1-z$) :
\begin{eqnarray}
\frac{8\,\pi^2\,\alpha\,\alpha_s\,C_F}{N_C\,Q \,z\,\bar z}
\Big(\frac{Q_u}{x-\xi+i\epsilon} + 
\frac{Q_d}{x+\xi-i\epsilon}  \Big)\epsilon^{n \varepsilon p \Delta}\!.\nonumber
\end{eqnarray}   

We choose $\phi_\rho(z)=6 z \bar z$   as the asymptotic
normalised meson  distribution amplitude and $f_\rho= 0.216$ GeV. After separating 
the real and imaginary parts of the amplitude, the $x$-integration gives:
\eqsal{
\label{eq:ivx}
\int_{-1}^1& dx\left(\frac{Q_u}{x-\xi+i\epsilon} + 
\frac{Q_d}{x+\xi-i\epsilon}  \right) V(x, \xi, t) \nonumber\\
=&Q_u \int^{1}_{-1} dx\,\frac{V(x,\xi,t)-
V(\xi,\xi,t)}{x-\xi}\nonumber\\
+&Q_d \int^{1}_{-1} dx\, \frac{V(x,\xi,t)-V(-\xi,\xi,t)}{x+\xi}\nonumber\\
+&Q_u V(\xi,\xi,t) (\log\left(\frac{1-\xi}{1+\xi}\right)-i\pi)\nonumber\\
+&Q_d V(-\xi,\xi,t)(\log\left(\frac{1+\xi}{1-\xi}\right)+i\pi ).\nonumber
}

The  scaling law for  the amplitude  is 
\begin{equation} 
{\cal M}^{TDA}_{\gamma^\star \gamma}(Q^2, \xi,t) \sim \frac{\alpha_s\sqrt{-t}}{Q}\;,
\label{scaling}
\end{equation} 
up to logarithmic corrections due to the anomalous dimension of the TDA.

\section{Models for TDAs}
To quantify the magnitudes of the cross section, we need to adopt a specific model for the non perturbative
TDA. 
As a first choice,  we start from a double distribution for $t=0$~\cite{rad}
which leads to a $x$- and $\xi$-dependence of the 
TDA  in the form $V^{(0)}(x,\xi)$:
\begin{equation}
\int_{-1}^{1} d\beta  \int_{-1+|\beta|}^{1-|\beta|}   d\alpha \; 
\delta(x-\beta -\xi \alpha)f(\beta,\alpha),\nonumber
\end{equation}
with $f(\beta,\alpha)= q(\beta)h(\beta,\alpha)$,
where $q(\beta)$ is analogous to the forward quark distribution in the 
GPD case and $h(\beta,\alpha)$ is a profile function  parametrised as
\begin{equation}
\frac{\Gamma(2b+2)}{2^{2b+1}\Gamma^2(b+1)}
\frac{[(1-|\beta|)^2-\alpha^2]^b}{(1-|\beta|)^{2b+1}}, \nonumber
\end{equation}
where the parameter $b$ characterises the strength of the $\xi-$dependence.
As a first guess, we assume that the $\beta-$dependence
of $q$ is given by a simple linear law 
$q(\beta) \;=\;2\;(1-\beta)\;\theta(\beta)\;$
and we assume a mild $\xi$ dependence as given by  $b=1$. Moreover
we implement the normalisation (with $\int dx G^{(0)}(x,\xi) = 1$) and the $t-$dependence of the TDA
through the vector form factor:
\begin{equation}
V(x,\xi,t) =V^{(0)}(x,\xi)\cdot \frac{f_\pi}{m_\pi}F_V(t)\;. \nonumber
\end{equation}
The $t-$dependence of this form factor has been studied in chiral perturbation theory  
and turned out to be weak, so we shall neglect it in this model and we shall use
the measured values at $t$=0.

\begin{figure}[h]
\includegraphics[width=7cm]{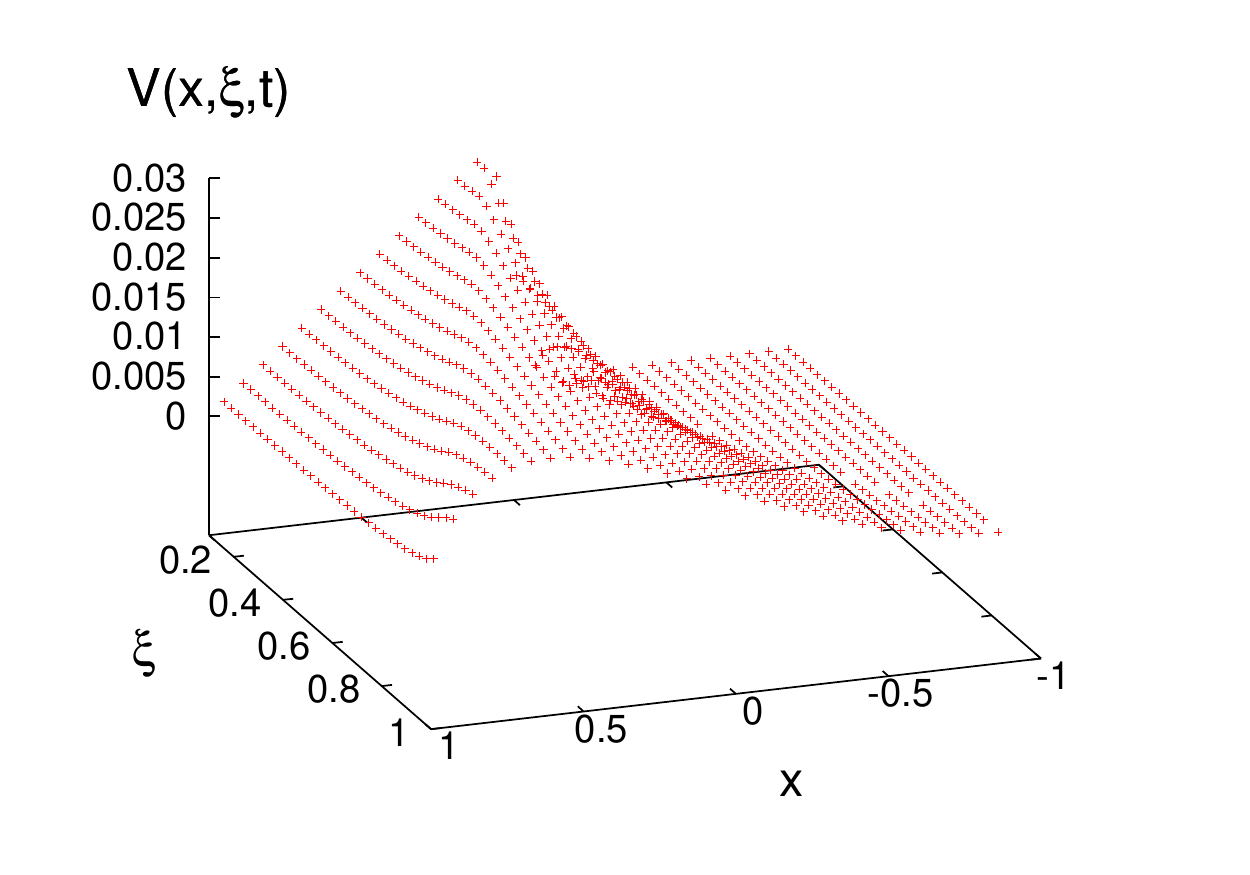}
\includegraphics[width=7cm]{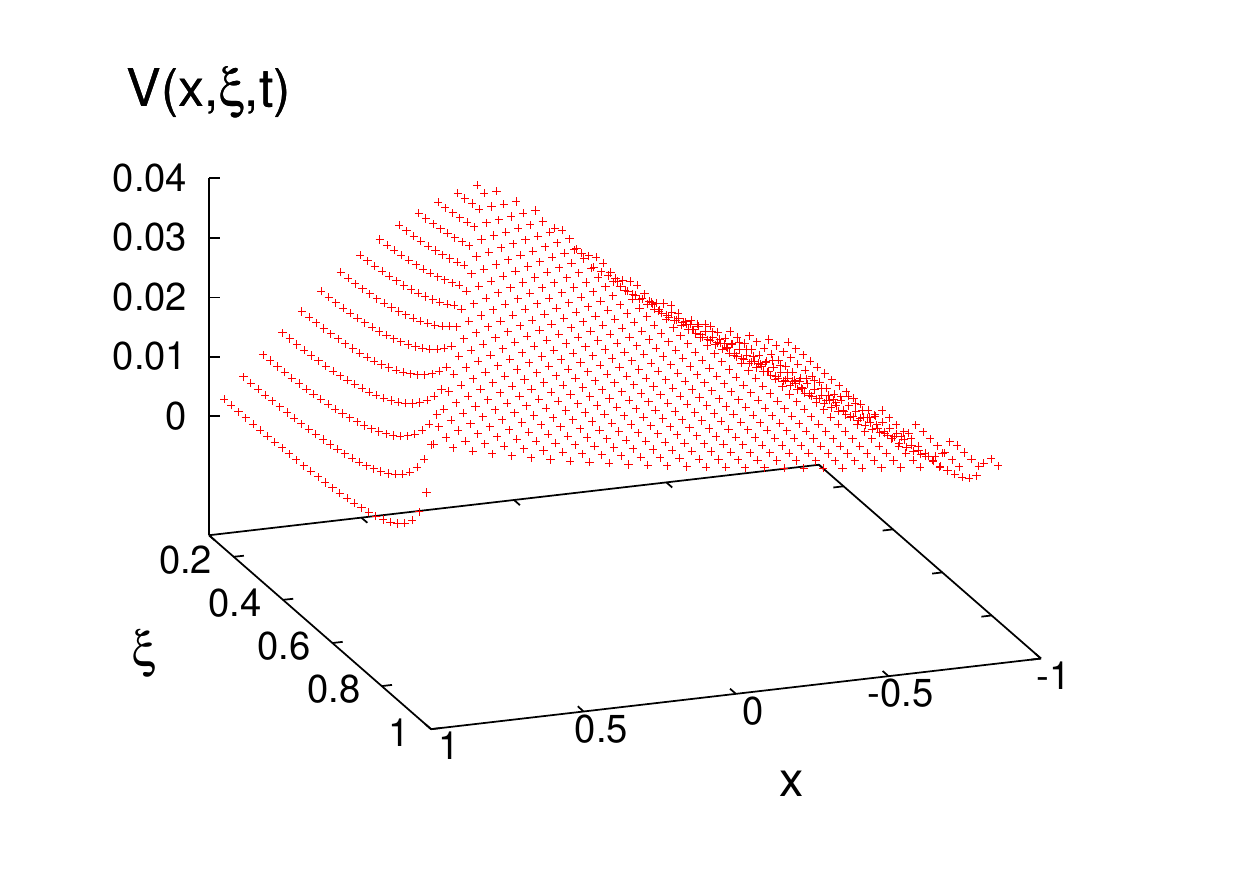}
\caption{The  $\gamma \to \pi^{-}$  vector transition distribution amplitude $V(x,\xi,t)$ in Model 1 
and in Model 2 (for $t=-0.5$ GeV$^2$).}
\label{GPITDA}
\end{figure}

As a second model, we  use  the initial $t$-dependent double distribution of Tiburzi~\cite{Tiburzi:2005nj}. 
Explicitly, $V(x,\xi,t)$ is written as
\eqsal{
& \frac{f_\pi}{m_\pi}\int_{-1}^{1}   d\beta  \int_{-1 + |\beta|}^{1 - |\beta|}
d\alpha \;
\delta(x - \beta - \xi \alpha)\times \nonumber\\&\frac{m^2}{4 \pi^2(m^2 + \frac{m_\pi^2 \beta ( \alpha + \beta - 1)}{2} -\frac{t(1 - \alpha^2 - \beta ( 2 - \beta))}{2})}\nonumber
}
where $m$ is set to $0.18$ GeV~\cite{Tiburzi:2005nj}. The sum rule  is satisfied 
for $t=-0.5$ GeV$^2$. We shall keep
$t$ fixed in the following to enable comparison with Model 1. In the following, 
we shall refer  to this approach as Model 2. We see that these two models give quite
 different results and we shall use both 
in section 3 to estimate the sensitivity of the cross sections to TDA models.
Other models for TDAs have recently been worked out \cite{Bro} and models
for pion GPDs~\cite{GPD_pion} could also be applied to the TDA case.

\section{Results and conclusions}

Since we want to focus on the study of the TDA behaviour, we decide to choose 
$Q^2$, $t$, $\xi$ and $\varphi$ as our kinematical variables.  
The differential cross section  reads 
\begin{eqnarray}
\frac{d\sigma^{e \gamma \to e \rho \,\pi}}{dQ^2 dt d\xi d\varphi}= \frac{|{\cal M}^{e \gamma \to e \rho \,\pi}|^2}{32 (2\pi)^4 s^2_{e\gamma}\xi (\xi+1)}.\nonumber
\end{eqnarray}

\begin{figure}[t!]
\centering\includegraphics[width=4.5cm,angle=-90,clip=true]{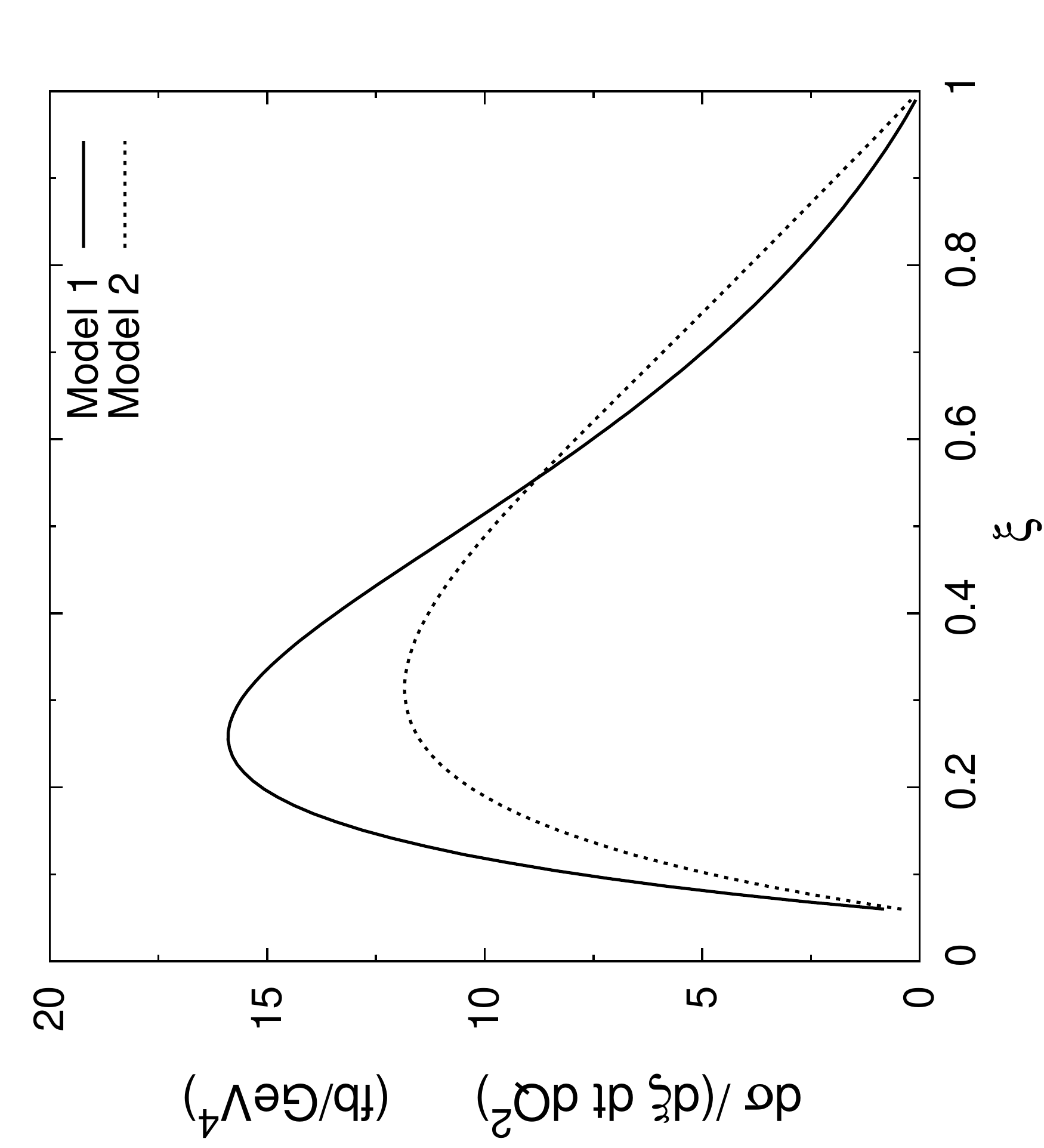}
\includegraphics[width=4.5cm,angle=-90]{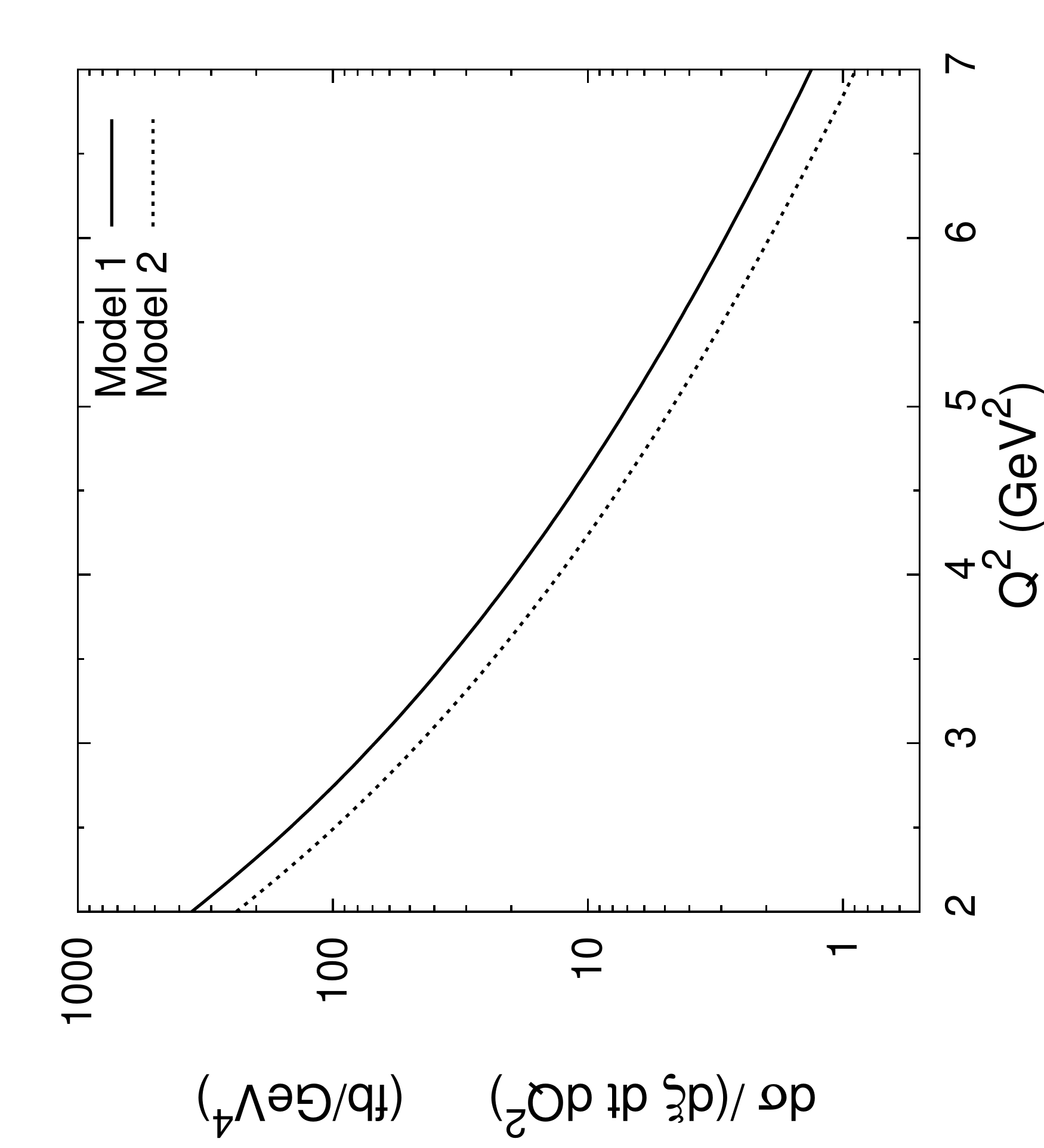}
\caption{$e\gamma \to e'\rho^+_L \pi^-$ differential cross section plotted as a function of
$\xi$ (a) and $Q^2$ (b) for $s_{e\gamma}=40$ GeV$^2$, $t=-0.5$ GeV$^2$ and respectively
$Q^2=4$ GeV$^2$ for (a) and $\xi=0.2$ for (b).}
\label{fig:rho_cross_section_xi_q2}
\end{figure}

In Fig.\ref{fig:rho_cross_section_xi_q2} (a),
we plot the cross section (after a trivial $\varphi$ integration)  vs $\xi$ and in 
Fig.\ref{fig:rho_cross_section_xi_q2} (b) vs $Q^2$ for both models of the vector TDA. 
The behaviour for intermediate values of 
$\xi$ is sensitive to specific models for the TDA. 
As shown in \ref{fig:rho_cross_section_xi_q2} (b), the $Q^2$-behaviour is model independent
and thus constitutes a crucial test of the validity of our approach.

Both real and imaginary part of the amplitude contribute significantly to the cross section,
which is reasonable at these moderate energies. Since the phenomenological analysis of the pion form factor
 indicates that a rather large  value of $\alpha_s$ should be used together with the asymptotic DA, we  use $\alpha_s=1$
in our numerical study.  Our  conclusions would not be strongly affected by a different choice.

It is now time to test   experimentally  the new factorised QCD approach to forward
hard exclusive scattering in $\gamma^\star \gamma$ exclusive reactions. We believe that our models for the photon to meson transition
distribution amplitudes are sufficiently constrained to give reasonable orders of magnitude for
the estimated cross sections. Cross sections are large enough for quantitative
studies to be performed. After verifying the scaling of the cross sections,
one should be able to measure these new hadronic matrix elements, and thus open a new gate to the understanding of the
hadronic structure.

%\verb?
%\section*{Acknowledgements}
%?

% ****************************************************************************
% BIBLIOGRAPHY AREA
% ****************************************************************************

\begin{footnotesize}
% IF YOU DO NOT USE BIBTEX, USE THE FOLLOWING SAMPLE SCHEME FOR THE REFERENCES
% ----------------------------------------------------------------------------

%****************************************************************
% END OF BIBLIOGRAPHY AREA
% ****************************************************************************
\end{footnotesize}
\end{document}